\documentstyle[11pt] {article}

\title{Time Arrow in Wave-Packet Evolution}
\author{R. Englman$^{a,b}$ and A. Yahalom$^{a,c}$ \\
$^a$ College of Judea and Samaria, Ariel 44284, Israel\\
$^b$ Department of Physics and Applied Mathematics,\\
Soreq NRC,Yavne 81810, Israel\\
$^c$ Faculty of Engineering, Tel-Aviv University,\\
Tel-Aviv 69978, Israel\\
e-mail: englman@vms.huji.ac.il, asya@ycariel.yosh.ac.il}

\begin{document}
\maketitle

\newcommand{\beq} {\begin{equation}}
\newcommand{\enq} {\end{equation}}
\newcommand{\ber} {\begin {eqnarray}}
\newcommand{\enr} {\end {eqnarray}}
\newcommand{\eq} {equation}
\newcommand{\eqs} {equations}
\newcommand {\er}[1] {equation (\ref{#1}) }

\begin {abstract}

	Availability of short, femtosecond laser pulses has recently made
feasible the probing of phases in an atomic or molecular wave-packet
(superposition of energy eigenstates). With short duration excitations the
initial form of the wave-packet is an essentially real "doorway state", and this
develops phases for each of its component amplitudes as it evolves. It is
suggested that these phases are hallmarks of a time arrow and irreversibility
that are inherent in the quantum mechanical processes of preparation and
evolution. To display the non-triviality of the result, we show under what
conditions it would not hold; to discuss its truth, we consider some apparent
contradictions. We propose that (in time-reversal invariant systems) the
preparation of "initially" complex wave-packets needs finite times to complete,
i.e., is not instantaneous.
\end {abstract}

\section {Introduction}

A quantum mechanical system developing in ordinary Hilbert Space
according to the time dependent Schr\"{o}dinger Equation does not exhibit the
time arrow. (Formal description of time-reversal invariance is found in almost
every text on quantum mechanics, e.g. [1], and is essentially implied by the
equality of the absolute magnitude of the wave function developing from zero
time at positive and at negative times.) On the other hand, the "time arrow'' is
manifest both in the subjective feeling of the passage of time and in statistical
mechanics involving macroscopic systems [2-4]. In some recent studies it was
therefore proposed to adopt a Rigged Hilbert Space, in which unidirectional
evolution, decay of an isolated elementary excitation and evolution of the
universe are all possible [5-6].
	We wish to focus on  "wave-packets", "doorway-states" and "observable
phases" and to show that, under circumstances that these are applicable, the
quantum mechanical evolution exhibits an experimentally verifiable time arrow
from the prepared state. The verifiability is made possible by recently
developed or perfected optical techniques for wave-packets of atomic
Rydberg states [7-8] and of molecular wave-packets [9]. (The quoted works
are recent samples of a multitude of relevant references over the past few
years.) In turn, these techniques are based on the possibility of experimental
observation of phases by use of coherent radiation (lasers) pulses of
femtosecond duration, which is the range appropriate to Rydberg-state and
molecular vibrational spectroscopies.

Let us examine the relationship between phase and time-direction in
the light of conventional treatments: The main sources for time-reversal
invariance (or its violation) are collisions between elementary particles [10]. In
that (quantum field theoretical) context, creation and destruction of particles is
allowed and time reversal (denoted by T) is normally considered in
conjunction with space inversion (P) and charge conjugation (C). The
invariance of the combination CPT is assumed in all interactions [11] while
based on experiments, CP or T is currently believed to be conserved in strong
and electromagnetic interactions, but slightly violated in weak interactions [12,
section 3.1]. In this work we shall use a single-particle (not field-theoretical)
description with a scalar wave-function that obeys a time dependent
Schr\"{o}dinger equation, and we assume that the Hamiltonian of the system is
time-reversal invariant. It then follows from the Schr\"{o}dinger equation (or from
the non-relativistic Lagrangian) that time reversal is equivalent to an inversion
of the phase [1]. This equivalence makes clear the relevance of recent phase
determinations to the question of time arrow.

        The essential features of the physical systems on which our result depends
are described in the next section. The whole issue is rephrased more
cautiously in the following section. Problems and questions that remain are
treated in the remainder.

\section {Short time preparation of the wave packet}

We start by considering (for the sake of simplicity) a time independent
Hamiltonian without a magnetic field. We add a sharp perturbation $\Delta H_{opt}$ that
causes a jump ("the preparation") from an initial stable state $|0>$, which can be
considered to be real. (For a dipole transition $\Delta H_{opt} = e \vec {\cal E} \cdot \vec r$,
 where ${\cal E}$ is a real electric field and $\vec r$ is the electron coordinate.)

The following is a typical scenario, due originally to Feshbach et al. for
scattering in nuclear matter [13], but later adopted and extended for
excitations into electron-vibrational states in molecules [14-18] Under
circumstances of moderately dense bunching of the discrete excited states
(such as would exist in excitation of Rydberg states [7] or of vibrational states
[9]) and a weak coupling to an adjacent continuum [14], the excitation is into a
doorway-state given by
\beq
|D> \propto \Delta H_{opt} |0>
\enq
When expanded into real energy-eigenstates, this can be written as
\beq
|D> = \sum_r a_r |r>
\enq
where (at $t=0$) all $a_r$ are real (except possibly for a common phase), since
$\Delta H_{opt}$ arises from a real field. The sum is over those energy eigenstates that
are included in the spectral width of the excitation. The physical condition for
the validity of the doorway description is that the duration of the excitation is
shorter than all relevant decay times. In this work, we shall call (for brevity)
preparation processes achieved under this condition "instantaneous" or
"occurring at $t=0$" (though, of course, the time-energy uncertainty principle
forbids state-selective excitations to be made instantaneously). Rhodes [16]
and Nitzan and Jortner [17] have derived equations (1) and (2) for excitation
times that are of the order of the inverse spectral width, which is the shortest
excitation time compatible with the time-energy uncertainty relation.

In contrast to the situation at $t=0$ when the $a_r$ are real, for $t>0$ the
coefficients become complex, with phases that differ for each component. The
precise formulae for $a_r (t)$ are given in [16-17] (and show these to be complex),
but for the simplest case, when the energy eigenstates are uncoupled to any
other state (discrete or in a continuum), the phases grow in time proportionally
to the energies of the r-states. We now come to an essential point in our
argument, namely, that this acquisition of phase is {\it irreversible} in the sense
that it is not possible to prepare with $\Delta H_{opt}$ a wave-packet whose coefficients
are complex at $t=0$ and become real at $t>0$. (We shall show this in more detail
in the sequel.) Moreover, the phases are observable (through coupling the
system to a phase-interferometer device, as was done, e.g., in [7]).

Also for $t>0$, when the r-states in the sum in (2) interact with
states outside the summation, certain coefficients decrease in an analogous
way to the behavior of a quasi-stable particle [6]. (For a system in which the
states outside the summation form a dense energy level scheme the
"decrease" takes a time-exponential form, but the decrease, subject to
Poincare recurrences only [18], exists also for microscopic systems in which
the r-states interact with discrete states outside the summation.) The
decrease is time-symmetric, since the absolute value of $a_r$ is even in $t$, but the
phase change of the coefficients is time {\it asymmetric}.

In summary, we have used a schematic description to argue that wave
packets prepared in a physical way (and moving in the full Hilbert space) are
by quantum mechanics time-asymmetric and irreversible.

\section {A Reformulation}

In the light of the foregoing we can restate our result, while phrasing
our argument somewhat more precisely.

We start with a suggested definition of reversibility.

A description may be said to be reversible, if the following state of
affairs holds: Whenever a "picture" (being a formal description of the physical
situation) $P_1$ at $t_1$ is followed by pictures $P_2$ at $t_2$,
$P_3$ at $t_3$,... $P_N$ at $t_N$, then the
description also allows that $P_N$ at $t_1$ will be followed by $P_{N-1}$ at $t_1 + t_{NN-1}$,
 ...$P_2$ at $t_1+t_{N2}$, $P_1$ at $t_1$+ $t_{N1}$ ($t_{rs}=t_r-t_s$), i.e. at the same
time intervals with the order of
events taken in the reverse sense. The times $t_r$ are arbitrary and not just some
times selected in a way that depends on the properties of the system.

[Two illustrations of the definition:

\noindent
(a) (A time-reversal invariant case) A classical particle, having
coordinate $q$, that moves under a constant force $f$ between times $t=t_1=0$ and $t_2$
is subject to the equation of motion: in suitable mass units $\frac{d^2 q}{dt^2} = a$.
This has a solution $q=\frac{1}{2} a t^2$, taking at $t_1$ the values: $q=0$,
$v=0$ (velocity), and at $t_2$: $q=\frac{1}{2} a t_2^2$, $v = a t_2$.
Starting at $t_2$ with a reverse velocity,
the solution is $q=\frac{1}{2} a (t-2 t_2)^2$, which has the following values
at a time $t$ which is
$t_2$ later (namely, $t=2 t_2$): q=0, v=0. This confirms reversibility for all choices of $t_2$.

\noindent
(b) (No time-reversal invariance). A particle moving in the $x-y$ plane
subject to a magnetic field $B$ perpendicular to the plane. The velocity
components satisfy $\frac{d v_x}{dt}=v_y B$, $\frac{d v_y}{dt}= -v_x B$,
with solutions $v_x=\sin {B t}$, $v_y=\cos{B t}$. The
values are at $t=0$: $v_x=0, v_y=1$ and at $t=t_2$: $v_x = \sin {B t_2}$, $v_y=\cos {B t_2}$.
The other solution commencing at $t=t_2$  is again $v_x=\sin {B t}$, $v_y = \cos {B t}$,
whose values at $t=t_2$ are identical to those of the previous solution.
However, at a time which is
$t_2$ later ($t=2 t_2$): $v_x=\sin {2 B t_2}$, $v_y = \cos {2 B t_2}$
and this can be made to agree with the
values of the first solution at $t=0$ only for special choices of $t_2$. This system
does not therefore satisfy our criterion for time-reversal invariance. By
changing the sign of $B$, as is done in a spin echo experiment [2], one gets the
solution $v_x=\sin {B (2 t_2-t)}$, $v_y = \cos {B (2 t_2-t)}$
and this indeed takes the values of the
original state at both times, but this involves a change in the system.]

We now return to the wave-packet. For definiteness, we take the case
that $P_1$ is the physical state (as represented by the many-component
wave-function in (2)) immediately after preparation of the wave-packet at $t_1=0$
by a short-duration excitation, and for that later pictures we take just one of
the several $P$'s, and call it $P_>$, the picture of the physical state at some
arbitrary later time $t_>$. Then we hope that the argument in the previous section
has shown that the picture reversal cannot hold, namely, $P_>$ cannot be
followed by $P_1$ at $t_>$, Thus the description in question (quantum mechanics,
including preparation and observation) is not reversible. The essential
difference is of course the absence of phases in $P_1$ and their presence in $P_>$.
The former occurs at $t=0$ and then only (under the mode of preparation here
envisaged), and the latter at $t>0$.

Have we proven a time arrow, or only a non-equivalence between the
time of preparation and time of observation?  If the sign of the phase is
observable (a proposition which we shall later put forward), then, given the
fact of lower-boundedness of energies, there is no question that measurement
of the phase establishes the time-arrow. However, even if the sign is
unknown, it is clear that the time direction of going from real-to-complex states
is uniquely aligned with earlier-to-later. This follows, since the observation of a
prepared state must take place at a time posterior to its preparation. Under
doorway-excitation conditions phases will be observed at later times than their
preparation-time and, indeed, one will get successively larger phases (at least
initially) as one carries out the observation at successively later times.

\section {Is the result trivial?}

We now raise the question, which ingredients in the description would have to
be removed, so as not to have a time arrow (or, to have "time-reversal
symmetry")? (We are now trying to show that our result is not trivial.)

\noindent
(1) If the phases were not observable, only the populations of the
components, then the pictures $P_1$ and $P_>$ would be identical and our
demonstration would fail. Though strictly this is true only for a finite number of
component states (energy eigenstates) that are not in overlap with a
continuum (recall our earlier remarks about the decrease of the absolute
value of $a_r$ when the r-states are coupled ), the crucial significance of phase is
already made clear.

\noindent
(2) In classical mechanics, one can start at $t=0$ with picture $P_1$ which could (as
an example) be the state that two particles which repel each other are placed
at some distance to each other and then get a picture $P_>$ at $t_>$ in which the
particles are at a farther distance from each other. However, one could also
start at $t=0$ with the picture $P_>$, impart to the particles velocities that are
opposites to those originally possessed by them at $t_>$ and then observe $P_1$ at
$t_>$. Thus in classical, phase-less mechanics our claim does not hold. (Note
also that, in the short-duration wave-packet preparation process from $|0>$, one
does not have the luxury of starting the process with negative phases: At $t=0$
there are just no relative phases of any sign. In the sequel we shall consider
a different preparation process, one which appears to be starting with
non-zero phases, but in fact does not do so.)

\noindent
(3) The previous parenthetical remark is invalid in a long-duration preparation
process or in preparations with one or more time delays. Then phases can be
introduced in any desired manner ("wave-function tailoring", [11]). If this
situation could be assumed to take place at t=0, then {\bf one could argue} that
one has started with $P_>$ and ended up at $t_>$ with $P_1$. (The bold faced phrase
"one could argue" is an expression of the possibility that the signs of the
phases cannot be measured. The bulk of present-day literature treats the
experimental determination of only the magnitude of phase-differences and
not their sign. The "observability" of phase is discussed by [20] and [21].
Recently, it has been shown that in certain circumstances the analytic
properties of the time-dependent wave-function can lead to the establishment
of a unique and unambiguously signed phase [22]. Based on a discussion
with Y. Aharonov, it appears that experimental determination of the sign of
phases is, in principle, possible.) It seems, however, that this situation ($P_>$
preceding $P_1$) cannot be regarded as the time-reversed counterpart of the
doorway-state excitation, since the preparation modes are essentially
different. Thus, the doorway-state preparation is "instantaneous", while the
creation of relative phases requires a finite time.

\noindent
(4) An alternative definition of time reversal invariance (one that, in our view,
would not do justice to the common understanding of the term), different from
that given at the beginning of this section, would not suffice to establish our
result. Thus, if time reversal invariance were defined as the possibility of the
reversal of pictures at some (rather than at any) later time, than the revival
time of a wave-packet [18], or a Poincare recurrence time, would constitute a
time starting from which phase growth could occur backwards.

\section {Apparent exceptions}

There are, however, cases of non-zero phases in a prepared state and these
should be noted. (We argue that they essentially differ from the "initial state",
as employed in the previous sections.)

\noindent
(a) On its face, it might have been possible to achieve a "phased" doorway
state by applying $\Delta H_{opt}$ onto the state $|0> +A |1>$, rather than as in (1) (Such a
"phase control, using a coherent laser source, was suggested some time ago
[23-24] and elaborated on in several works more recently [7-8].) The
coefficient $A$ can have any desired magnitude and phase, through choices of
the laser intensity and of the time delay between the applications of the laser
on $|0>$ and the later application of $\Delta H_{opt}$ on $|0> +A |1>$.
 However, the "phased doorway state" obtained in the aforesaid
manner is not a "prepared" state in our sense, since it does no arise at $t=0$
(which is defined as the moment of preparation) but only after a time delay.
The "time-arrow test by exploration of the phases" is to be conducted at the
moment of preparation ($t=0$) and at a moment of observation, which is
different from $t=0$, rather than at two times different from zero. Moreover, (not
only the preparation of a complex wave-packet, but also) the observational
check that a wave-packet state has complex coefficients requires a prior
procedure (through the employment of appropriate time delays) that takes
place some time before the moment of observation on the wave packet. In
other words, no instantaneous checking of phases is possible, either.

\noindent
(b) A broad class of wave packets that have complex coefficients have been
given by a large number of workers.(E.g., [25,26]) These states have a
non-zero initial momentum. In the next section ("An application") it will be
shown that to prepare such a complex wave packet, one needs an excitation
that has a finite time duration, or a time dependent coupling mechanism.

\noindent
(c) A free particle state of the form $e^{ikx}$ is necessarily complex. (In other words,
the amplitudes of position eigenstates are complex in a momentum
eigenstate.) Its special status may be due to its not being a bound state, and
indeed if the free particle is confined to a box with reflecting (or absorbing)
boundaries, then the amplitudes become real.

\noindent
(d) Complex states $e^{im \theta}$ (where $2m$ is an integer and $\theta$ an angle)
can arise with magnetic excitation. The preparation involves a Hamiltonian that is not
time-reversal invariant and is excluded from our present considerations.

\section{An application}

In the field of molecular wave-packet dynamics, complex  "initial" wave-packet
states are frequently employed [25,26]. A preparation process for a diatomic
molecule has been recently described with great clarity and with provision of
considerable formal details [9]. Here we give a simplified and condensed
description, using a wave-function rather than a density-matrix formalism. The
essential point for the present purpose is the non-instantaneous preparation
of a complex state and this is clearly noted also in [9]. The accompanying
drawing (Figure 1) is a schematic version of Figure 2 in [9].

\begin{figure}
\vspace{4cm}
\begin{picture}(1,1)
\end{picture}
\includegraphics{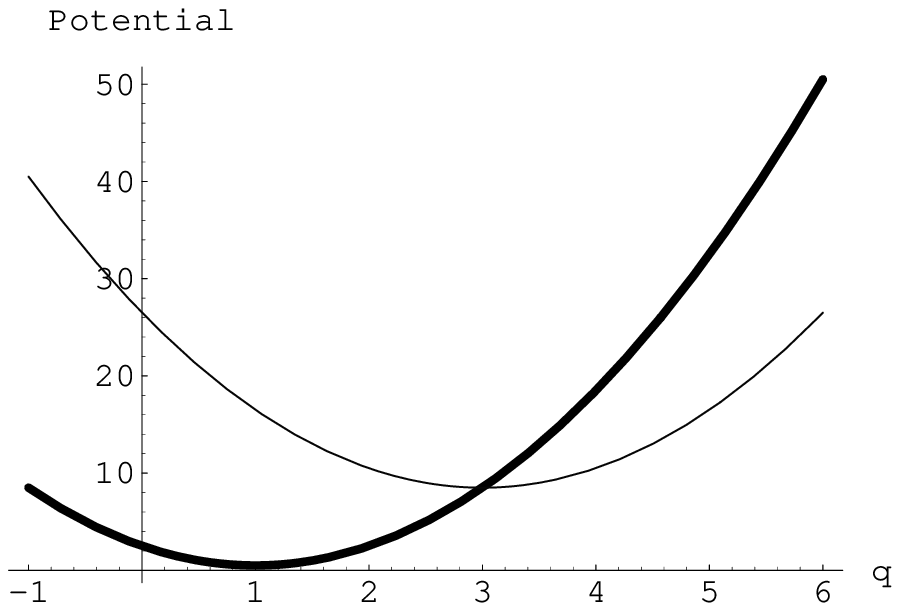}
\caption {Schematic representation of two potential energy surfaces (in arbitrary
units) belonging to two electronic states, as function of an atomic coordinate
$q$. A short duration electromagnetic excitation from the ground level of the
lower potential raises the system to the upper potential, where the wave
packet is formed.}
\label {fig: nad1}
\end{figure}

We consider a wave packet, which is prepared in the form:
\beq
\Psi (q) = {\cal N} e^{ikq} e^{-M \frac{\omega}{2}(q-q_o)^2}                     					
\enq
Here $q$ is the coordinate of a particle (which could be, say, an atom of mass $M$
constrained to move inside a parabolic potential wells centered in $q_o$ in the
electronic ground state, and inside a parabolic potential centered at some
other point in an excited electronic state, to which it is raised by an
electromagnetic field), $\omega$ is the frequency of harmonic motion, ${\cal N}$ is a
normalizing factor and $k$ is the initial momentum
of the particle. ($\hbar =1$.) The
complex form of the initial state is apparent in (3). In the sudden
approximation for the excitation process, its representation by the (real)
eigenstates $\phi_n (q)$ of the upper potential is
\beq
\Psi(q) =\sum_n (a_n +i b_n) \phi_n (q)             					
\enq
where $a_n=<Re \Psi(q)|\phi_n (q)>$, $b_n =<Im \Psi(q)|\phi_n(q)>$

A physical process which ensures the creation of the wave packet (3) or (4) is
the application in the ground (electronic and vibrational) state of several
coherent (real) electric fields. It was shown in [24] that when a sharply pulsed
laser light is applied, each coefficient $a_n$ or $b_n$, though initially time varying, will
settle down to its long-time value after a time exceeding the pulse duration
(which may be as short as several femtoseconds). We represent the real field
(during the pulse) by
\beq
{\cal E} (\Omega)= {\cal E}_0 \cos (\Omega t)								
\enq
whose oscillatory frequency is  $\Omega$. The field creates a nonzero transition
dipole between the ground and the excited electronic state. To obtain the
wave packet (4) at, say, $t_0$ we need to apply a set of fields some of which, it
turns out, have to be advanced in time. First, to construct the $a_n$ part of the
wave packet, we apply at $t=t_0$ the superposition
\beq
{\cal E}_1 = \sum_n (\frac{a_n}{F_n}) {\cal E} (n \omega + \Delta)
\enq
Next, to obtain the imaginary $b_n$ terms we apply at $t= t_0 - \delta t_n$,
a further set of fields consisting of the sum of fields
\beq
{\cal E}_2 = \sum_n (\frac{b_n}{F_n}) {\cal E} (n \omega + \Delta)
\enq
In the above $\Delta$ is the energy gap between the two electronic states, $F_n$ is the
Franck-Condon amplitude (=square-root of the F-C factor) for transitions
between the 0 and a displaced n'th vibrational level. (For F-C factors see [15,
27].) The time-advance $\delta t_n$ for the excitation time in  (7)
is such that later, at $t_0$,
the phase factor of the excited level at the height $n \omega + \Delta$ is just
\beq
e^{i\frac{\pi}{2}} = i       									
\enq
and this is achieved by so choosing $\delta t_n$ that
\beq
(n \omega + \Delta) \delta t_n= \frac{\pi}{2} mod(2 \pi)
\enq
 The required time advances need to be longer than the pulse width, and this
is feasible with femtosecond pulses. Also, there is no problem, in principle, of
letting light from a coherent source arrive at different times and intensities,
employing path deflection into dielectric and partly absorbing channels.

It is now elementary to check that the "initial" state (i.e., the state at $t_0$)
in  (4) is reconstructed, but only by starting the preparation process earlier
than $t_0$. This example illustrates our earlier statement that the preparation of
a complex doorway state cannot be achieved instantaneously.

\section {A paradox}

A nutshell summary of our results might be:
One-time preparation means no time reversal; multiple-time preparation can
exhibit time reversed development.

It may seem odd that touching the system just once gives a
time-direction, whereas manipulating it at subsequent times restores the time
reversal invariance.

However, classical systems show just this behavior, as we shall now
illustrate. Suppose a (classical) particle is set into motion at $t=0$ in a Universe
that is such that the masses of particles decrease by a fixed fraction after
each second. Obviously this Universe contains a time-arrow (just as our
Quantum Universe does). The position of the "kicked" particle is shown in Fig.
2 as function of time up to $5$ seconds with the upward curve, (This is a sum of
five straight line fragments, with successively increasing slopes.) Trying to
obtain a time-reversed motion, we cannot just change the sign of the velocity
which was present at $5$ seconds and let go, as is shown by the downward thin
line. Rather, we must interfere at the tick of each second to adjust the speed
(downward thick lines in Fig. 2). Thus, in a classical system, too, time reversal
can be got artificially, so to speak, by repeated manipulation.

\begin{figure}
\vspace{4cm}
\begin{picture}(1,1)
\end{picture}
\includegraphics{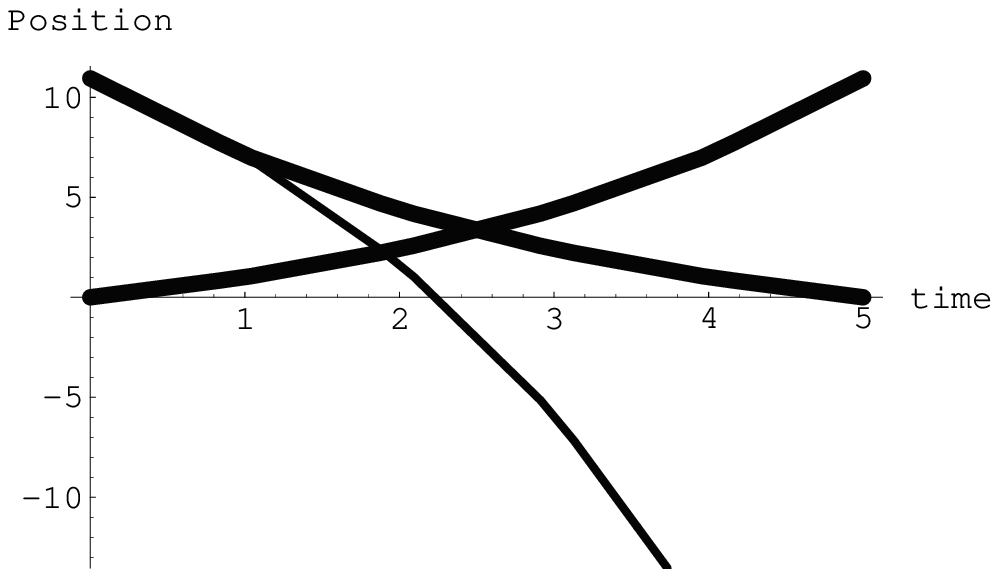}
\caption {Position (in velocity-seconds unit) as function of time (in seconds), of a
classical particle whose mass decreases by a fixed fraction $(\frac{1}{1.4})$ after each
second, plotted for the first five seconds.
The upward curve represents free motion with an initial positive velocity.
The thin downward curve shows the position of the same particle if this starts
from its position after five seconds and is given a velocity which is the
opposite to that which the particle had at that instant, and is then left moving.
The motion is {\bf not} the time-reversed one of the previous.
The thick downward curve shows the position, if the particle velocity is
{\bf adjusted after each second} so as to have the opposite velocity to that in the
upward curve. The motion is now the time-reversed one to that in the upward
curve.}
\label {fig: 2}
\end{figure}

\section {Future directions}

The following points need further consideration:
Is time-asymmetry a special phenomenon that occurs only in the
doorway-state spreading, or do asymmetry and irreversibility occur more
generally, under any state-preparation procedure. If the latter is the case, how
can it be demonstrated? In particular, the relation to time arrows in particle
physics and involving basic forces (but exclusive of the gravitational one)
need to be discussed. As already noted, our result depends on the practical
feasibility of short time pulsed preparation of the initial wave-packet state (also
known as the sudden approximation). In collision or scattering of nucleons
and elementary particles the preparation of the incoming state normally starts
at $t= -\infty$ and takes place adiabatically. The difference in the two situations  for
fundamental processes was emphasized recently [28]. A suggestion to apply
at $t=0$ a 'pulse' of particles to bombard a stable nucleus was made some time
ago by Khalfin [29]. We are not aware of any phase or other measurements
that arose from the suggested experiments. More recently, the theory of T.D.
Lee [30] about spontaneous T-violations through phase interference has
received renewed attention with a view to experimental tests [31,32]. These
ideas are clearly beyond the scope of the present paper,  which deals with
established technology.

How is the above "molecular" irreversibility, occurring after preparation,
connected with either the statistical-mechanical  time arrow, or with the
irreversibility entailed by a measurement process, or with the irreversible
processes considered in [5-6] (decay of elementary excitations and evolution
of the Universe)?

The foregoing theory has to be fitted in with the description, originally
due to Aharonov and coworkers [33], of retrodiction from a performed
measurement.
		
\section{Experiments}

Recent experiments on Rydberg-states appear to confirm the
existence of a "doorway state", but one of the wave-packets in the
experiments of [7] appears to have one non-zero phase (of $.15 \pi$), in apparent
opposition to the theory. The seven other components have real coefficients
as required by the foregoing theory. It appears that the complex coefficient is
due either to a time-advanced mode of preparation of the wave packet or to
the phase determination being made at some later time than the creation of
the wave packet.

It has also been remarked to us that if the preparation and
(subsequent) observation procedures are "put in", then the time arrow must
necessarily "come out", since the "extended" Hamiltonian is time-dependent
(rather than time-reversal invariant, as is the Hamiltonian of the measured
system). This, however, only means that the formalism is consistent upon
extension of what is meant by a "system". The essential point remains that the
wave function of the measured system is observably different at an initial time
(the moment of preparation) and at a later time (the moment of observation),
with the difference showing up through the relative phases in the component
amplitudes.

\section{Conclusion}

It has been shown that when a wave-packet is prepared in currently
practized ways, such as described in [7-9], its "properly initial" and its
later-time states differ by the acquisition of phases, which are odd in $t$ (time).
Recently developed techniques for atomic and molecular wave-packets can
test for phases and, by consequence, for time direction.  Even if it is doubted
that phase signs are observable, in wave-packets that are prepared via
"doorway-state" excitation the weaker, but certainly verifiable reality -
complexity property of amplitudes is uni-directionally aligned with a time-arrow
implied by the preparation- observation procedure [6].

A further proposition of this work, supported with examples, is that, in
systems with time-reversal invariant and spatially bound Hamiltonians,
complex (=phased) wave-functions cannot be prepared (or measured)
instantaneously.

Several issues have been relegated for future work. This ought to show
how the results of this paper will appear when translated from the specific
setting of atomic and molecular wave-packets to more general ones.

\section {Acknowledgments}

Thanks are due to Yakir Aharonov for clarifications and to A. Elitzur, M.E.
Perel'man and M. Shapiro for discussions.

\end{document}